\documentclass[useAMS,usenatbib,referee]{mn2e}
\usepackage[pagewise]{lineno}
\usepackage{url}
\usepackage{color}
\usepackage{microtype}
\usepackage[colorlinks=true,citecolor=blue]{hyperref}
\usepackage{graphicx,graphics,color}

\newcommand\chandra{{\it Chandra}}
\newcommand\rosat{{\it ROSAT}}

\newcommand\xmm{{\it XMM-Newton}}

\newcommand\ks{{\rm~ks}}

\newcommand\kev{{\rm~keV}}
\newcommand\ev{{\rm~eV}}
\newcommand{\degree}{$^{\circ}$}
\newcommand\kms{\ifmmode {\rm~km\ s}^{-1} \else ~km s$^{-1}$\fi}
\newcommand\Hunit{\ifmmode {\rm~km\ s}^{-1}\ {\rm Mpc}^{-1}
        \else ~km s$^{-1}$ Mpc$^{-1}$\fi}
\newcommand\ctssec{\ifmmode {\rm~count\ s}^{-1} \else ~count s$^{-1}$\fi}
\newcommand\ergsec{\ifmmode {\rm~erg\ s}^{-1} \else
        ~erg s$^{-1}$\fi}
\newcommand\funit{\ifmmode {\rm~erg\ s}^{-1}\;{\rm cm}^{-2} \else
        ~ergs s$^{-1}$ cm$^{-2}$\fi}
\newcommand\phflux{\ifmmode {\rm~photon\ s}^{-1}\;{\rm cm}^{-2}
        \else   ~photon s$^{-1}$ cm$^{-2}$\fi}
\newcommand\efluxA{\ifmmode {\rm~erg\ s}^{-1}\;{\rm cm}^{-2}\;{\rm
        \AA}^{-1} \else ~erg s$^{-1}$ cm$^{-2}$ \AA$^{-1}$\fi}
\newcommand\efluxHz{\ifmmode {\rm~erg\ s}^{-1}\;{\rm cm}^{-2}\;{\rm
        Hz}^{-1} \else ~erg s$^{-1}$ cm$^{-2}$ Hz$^{-1}$\fi}
\newcommand\cc{\ifmmode {\rm~cm}^{-3} \else cm$^{-3}$\fi}
\newcommand\FWHM{\ifmmode {\rm~FWHM} \else ${\rm~FWHM}$\fi}
\newcommand\Msun{\ifmmode M_{\odot} \else $M_{\odot}$\fi}
\newcommand\Lsun{\ifmmode L_{\odot} \else $L_{\odot}$\fi}

\newcommand\hbeta{\ifmmode {\rm H}\beta \else H$\beta$\fi}
\newcommand\Kalpha{\ifmmode {\rm K}\alpha \else K$\alpha$\fi}
\newcommand\nh{\ifmmode N_{\rm H} \else N$_{\rm H}$\fi}
\newcommand{\mnras}{MNRAS}

\title[CTD~86]{A Multi-wavelength study of nuclear activity and environment of
low power radio galaxy CTD~86}
\author[Pandge et al.] {M. B. Pandge$^{1}$, G. C.
Dewangan$^{2}$, K. P. Singh$^{3}$, M. K. Patil$^{1}$\thanks{E-mail:
patil@iucaa.ernet.in} \\
$^{1}$Swami Ramanand Teerth Marathwada University Nanded 431606, India\\
$^{2}$Inter-University Centre for Astronomy and Astrophysics, Post Bag 4,
Ganeshkhind,  Pune 411007, India \\
$^{3}$Tata Institute of Fundamental Research, Mumbai 400005, India.\\
}

\begin{document}



\maketitle

\label{firstpage}

\begin{abstract}
We present an X-ray study of the nuclear and extended emission of a  nearby
Fanaroff \& Riley class I (FR-I) radio galaxy CTD~86 based on the \xmm{}
observations.  Two different components observed are : diffuse thermal emission
from hot gas ($kT\sim  0.79\kev$, $n_e\sim
10^{-3}{\rm~cm^{-3}}$, $L_X \sim  5\times10^{42}{\rm~erg~s^{-1}}$ extended over
$\sim 186{\rm~kpc}$), and unresolved nuclear emission exhibiting mild
activity.  The hot gaseous environment of CTD~86 is similar to that found in
groups of galaxies or in bright early-type galaxies.  No clear signatures of
radio-lobe interaction with the diffuse hot gas is evident in this case.  X-ray
emission from the nucleus is well constrained by an intrinsically absorbed ($N_H
\sim 5.9\times10^{22}{\rm~cm^{-2}}$) power law ($\Gamma \sim 1.5$) with
$2-10\kev$ luminosity $L_X \sim  2.1\times10^{42}{\rm~erg~s^{-1}}$.  We have
measured the stellar velocity dispersion, $\sigma=182\pm8\kms$, for the CTD~86
and  estimated a mass $M_{BH}\sim 9\times 10^7{\rm~M_\odot}$ 
with $L_{bol}/L_{Edd} \sim 4\times10^{-3}$.  The low $L_{bol}/L_{Edd}$ rate and
high $L_X/L_{[O~III]}$ ratio suggest that the central engine of CTD~86 consists
of a truncated accretion disk lacking a strong ionizing UV radiation and an
inner hot flow producing the X-ray emission. The truncated disk is likely to be
inclined with ($i\sim40^\circ-50^\circ$) such that our line of sight passes
through the outer regions of a putative torus and thus results in high X-ray
absorption.  We have also identified two bright X-ray sources,
SDSS~J142452.11+263715.1 and
SDSS~J142443.78+263616.2, near CTD~86.  SDSS~J142452.11+263715.1 is a type 1
active galactic nucleus at $z=0.3761$ and unabsorbed $0.3-10\kev$ X-ray
luminosity $L_X\sim 8 \times 10^{43}{\rm~erg~s^{-1}}$, while SDSS
J142443.78+263616.2 is probably a galaxy with an active nucleus.
\end{abstract}


\section{Introduction}
Optical and X-ray luminosity functions of active galactic nuclei (AGN) are not
complete at the faint end of luminosities.  Both radio-quiet as well as
radio-loud active galaxies are expected to exist at low optical and X-ray
luminosities.  The nearby low luminosity radio galaxies (LLRGs) are good
candidates to study the mild nuclear activity in galaxies.  Since LLRGs also
represent the parent population of BL Lac objects, therefore, their study can
provide an insight into the unification and evolution of radio galaxies.  The
LLRGs are members of Fanaroff-Riley class 1 (FR-I) \citep{1974MNRAS.167P..31F},
 where one finds a variety of sources in terms of power and radio
morphology. Apart from having radio power $\leq 5 \times10^{25} {\rm~W~Hz^{-1}}$
at a frequency of 1.4 GHz, FR-I sources have only one feature in common: hot
spots at the outer edge of lobes are never seen.  Naked radio jets occur
preferably at the very low end of radio power ($< 10^{23} {\rm W~Hz^{-1}}$), 
while at the upper end ($\sim10^{25}{\rm~W~Hz^{-1}}$) one sided jets and hot
spots in the middle of the lobes are seen.  The radio emission is thought to
arise from the synchrotron radiation mechanism and is a manifestation of nuclear
activity.

The nuclear activity is also inferred from other independent means such as
optical and X-ray emission. Supermassive black holes, accretion discs, broad
line regions (BLRs), narrow line regions (NLRs) etc., all are thought to be
associated with low luminosity radio galaxies (LLRGs).  Therefore, X-ray and
optical studies of LLRGs are useful to estimate some of the basic parameters, e.
g., black hole mass, accretion rate, velocity of BLRs and NLRs etc., governing
the overall appearance of these galaxies.   \citet{2004ApJ...617..915D} found
central compact X-ray cores in 13 systems out of 25 FR-I radio galaxies from the
3CRR (Third Cambridge Catalogue of Radio Sources) \citep{1985PASP...97..932S}
and B2 (Second Bologna Catalog of radio sources)
\citep{1975A&AS...20....1C,1978A&AS...34..341F} catalogs. From flux$-$flux and
luminosity$-$luminosity core X-ray/radio correlations 
\citet{1999MNRAS.310...30C} and \citet{2004ApJ...617..915D} found a physical
relationship between the soft X-ray emission of radio galaxies and the
jet-generated radio core emission, suggesting that at least some of the X-ray
emission is related to the nuclear jet.  \citet{2004ApJ...617..915D} also
reported small Eddington ratios, $L_{bol}/L_{Edd} \sim 10^{-3} - 10^{-8}$ and
failed to confirm strong X-ray absorption; thus implying that their sample of
FR-I galaxies generally lack a standard torus.  Since the Eddington ratios of
FR-I galaxies are very small compared to those of Seyfert 1 galaxies and
quasars, it is possible that the central engine of FR~I galaxies is different
from that of other AGN with high Eddington ratios.  Therefore, multiwavelength
study of central compact core of FR-I galaxies is required to make any progress
in our understanding of their central engines.

The massive elliptical host galaxies, many of which contain radio-loud AGN, 
are associated with large amount of hot gas with temperatures of
$10^6-10^8{\rm~K}$  
\citep{1986RvMP...58....1S, 2000ARA&A..38..289M,
2003ARA&A..41..191M, 2005RSPTA.363.2711H}.  The gas is heated in the deep
gravitation potential wells of the massive elliptical galaxies or the associated
groups or clusters of galaxies.  It is now well known
that the radio lobes and jets of radio-loud AGN interact with the
environment and can efficiently heat the surrounding medium and thus prevent the
accumulation of large amount of cool gas in the central regions \citep[][and
references therein]{2005RSPTA.363.2711H}.  

Here, we present results based on the X-ray observations (\xmm,
\citealt[][]{2001A&A...365L...1J}) and compare it with the optical (SDSS, Sloan
Digital Sky Survey) and radio (FIRST, Faint Images of the Radio Sky at Twenty
Centimeters; \citealt{1994ASPC...61..165B}) data of the FR-I radio galaxy CTD~86
(CalTech list D of radio sources), also known as (B2~1422+26).  CTD~86 is a
low-power radio galaxy \citep{1987A&A...181..244P,1999MNRAS.310...30C} with
radio power 1.9$\times10^{24}{\rm~W Hz^{-1}}$ at 1.4 GHz, photometric B band
magnitude of 15.62 and is located at a distance of about 161 Mpc
(H$_0$ =73 ${\rm~km~s^{-1}} $Mp$c^{-1}$, z=0.037).  CTD~86 is an elliptical
(E2-type) radio galaxy and belongs to the poor cluster AWM~3
\citep{1984ApJ...283...33B} with $\sim$44
members at a redshift of $cz=4495{\rm~km~s^{-1}}$.  The global
parameters of CTD~86 are summarized in Table \ref{basicpro}.  The radio
source B2 1422+26B (observed with the Effelsberg 100-m telescope
\cite[]{1994A&AS..103..157M}) associated with CTD~86 shows a symmetric double
lobed structure and is extended up to about 75 kpc from the optical center of
the galaxy.  At the outer edges of the lobes, symmetrical ring like
structures are also evident.  The structure of this paper is as follows.
Section 2 describes the X-ray data analysis and presents X-ray images and
spectra obtained. Section 3 provides a discussion on the results, and is
followed by conclusions in Section 4.

\begin{table*}
  \caption{Global parameters of CTD~86}
  \begin{tabular}{@{}llrrrrlrlr@{}}
    \hline
    \hline
    $\alpha$ (J2000); $\delta$ (J2000)$^{1}$ &14:24:40.5; +26:37:31 \\
     Morphological type$^{2}$ & E2  \\
    Magnitude$^{3}$ ($B$) & 15.62\\
    Size$^{1}$ & 0\arcmin.96$\times$0\arcmin.66\\
    Distance$^{3}$ (Mpc) & 161\\
    $z^{4}$ & 0.037152\\
    Radio core flux density$^5$
    ($f_{5GHz}$) & 25mJy \\
    \hline
    \hline
  \end{tabular}

  $^1$ {NED: NASA/IPAC  EXTRA GALACTIC  DATABASE }; $^2$
\cite{1968AuJPh..21..903M}; $^3$
\citep{1987A&A...181..244P,1999MNRAS.310...30C}; $^4$
\cite{2002AJ....123.3018M} $^5$ \cite{1991A&A...252..528G}

  \label{basicpro}
\end{table*}

\begin{figure}
  \centering
  \includegraphics[scale=0.45]{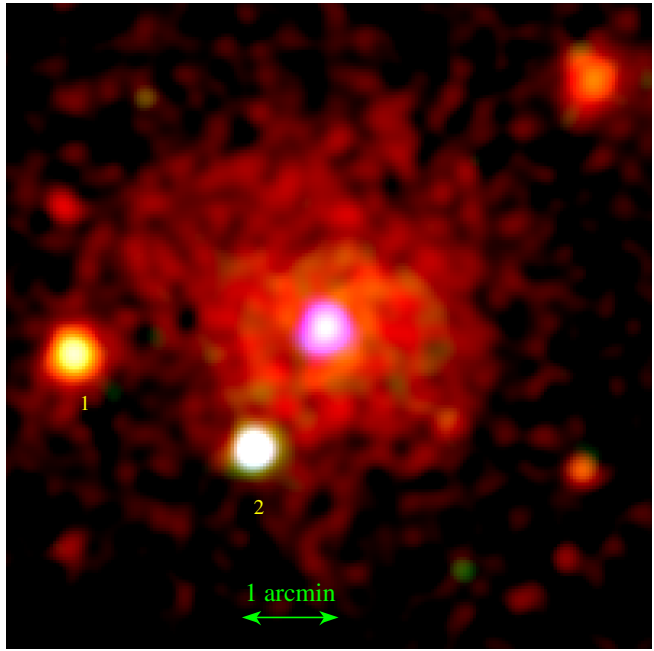}
  \caption{\xmm{} tri-color image of CTD~86 and its environment.  The
    soft ($0.3-1\kev$) band is shown in red, the intermediate
    ($1-2\kev$) band in green and the hard ($2-10\kev$) band is shown
    in blue. One arcmin corresponds to a linear size of
    $43.7{\rm~kpc}$ at the red shift of CTD~86 ($z=0.037216$). The
    unresolved source marked ``1" is SDSS~142452.11+263715.1 and the
    source ``2" is SDSS~J142443.78+263616.2. Source ``1" shows broad
    emission lines in its optical spectrum and is classified as a
    quasar at $z=0.3761$. Source ``2" is also most likely an AGN.}
  \label{xray_color}
\end{figure}

\begin{figure}
  \includegraphics[scale=0.45]{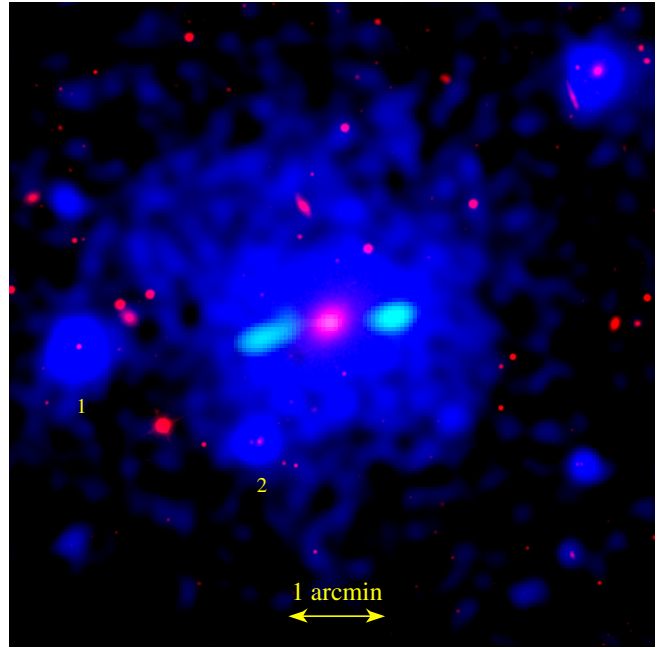}
\centering
  \caption{Composite soft ($0.3-1\kev$) band X-ray (blue), SDSS
    optical (red) and FIRST radio (green) image of CTD~86 and its
    environment.  The unresolved source marked ``1" is
    SDSS~142452.11+263715.1 and the source ``2" is
    SDSS~J142443.78+263616.2.  Optical counterparts of the point
    sources are clearly evident.}
  \label{xray_opt_radio}
\end{figure}

\begin{figure}
  \centering
  \includegraphics[width=100mm]{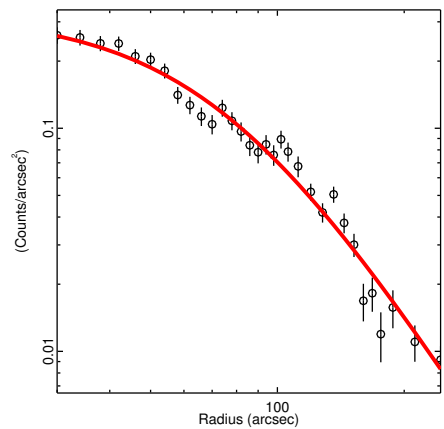}
  \caption{The azimuthally averaged, background subtracted radial surface
brightness profile of CTD 86 in the $0.3-2.5\kev$ energy band. The
red continuous line shows the best fitted 1d-$\beta$ model and black points are
the extracted photons from each of the annulus.}
  \label{sur}
\end{figure}

\section{The \xmm{} data}
CTD~86 was observed twice with \xmm{}, on 2010 June 21 (obsid: 0652720101;
hereafter Obs.I) and  2010 August 02 (obsid: 0652720201; hereafter Obs.II),
with effective exposure times of  $33.9\ks$ and $32.4\ks$, respectively.  In
both the observations, all the three EPIC cameras were operated in the full
frame mode using the ``thin'' filter for the EPIC-pn \citep{2001A&A...365L..18S}
and the ``medium'' filter for the two MOS cameras \citep{2001A&A...365L..27T}.
We have made use of both the data sets on CTD~86 available in the archive of the
HEASARC.  The observation data files (ODF) for CTD~86 obtained from the \xmm{}
archive, were processed using the Science Analysis Software (SAS) version 11.
Examination of the background rate above $10\kev$ revealed that the data were
partly affected by the flaring particle background.  Time intervals with the
count rates greater than $0.6{\rm~counts~s^{-1}}$ for the EPIC-pn and
$0.175{\rm~counts~s^{-1}}$ for MOS1 and MOS2 detectors were identified as
background flares and were neglected during further analysis.  Only events
corresponding to patterns 0--4 for the EPIC-pn and 0--12 for the MOS detectors
were retained and used for further analysis.  The resulting net EPIC-pn, MOS1
and MOS2 exposure times for Obs.I are 30 ks, 30.1 ks and 25 ks and for Obs.II
are 27.81 ks, 27.12 ks, 27.12 ks, respectively.

We have a composite (PN+MOS1+MOS2) mosaic image from the Obs.I dataset by
combining the EPIC-pn, MOS1 and MOS2 event files in the soft ($0.3-1\kev$),
intermediate ($1-2\kev$) and hard ($2-10\kev$) bands.  These images were
smoothed with a Gaussian kernel of radius $10\arcsec$ and were then combined to
from a single image in color coded form Figure~\ref{xray_color}.  This figure
shows the tri-color X-ray image of CTD~86 and its environment and confirms the
presence of diffuse soft X-ray emission surrounding CTD~86 reported previously
by \cite{1999MNRAS.310...30C}.  Two unresolved X-ray sources near
CTD~86 are also evident in this figure and were reported previously
with \rosat{} \cite{1999MNRAS.310...30C}.  In addition, this figure revealed an
unresolved source, prominent in the ($2.0-10.0)\kev$ band, in the core of
CTD~86. We also made a composite X-ray ($0.3-1.0)\kev$ band (blue), SDSS
optical (red) and radio FIRST (green) image for CTD~86 and is shown in
Figure~\ref{xray_opt_radio}.  This figure clearly shows an unresolved X-ray core
at the center of the CTD~86. Thus, hard X-rays from the central part of overall
X-ray emission most likely represents the active nucleus of CTD~86.  The
radio-lobes of CTD~86 are embedded in the diffuse X-ray emission surrounding the
nucleus.  Our analysis failed to detect X-ray cavities formed due to the
interaction of radio-lobes with the surrounding hot gas
\citep{2012MNRAS.421..808P}.

\subsection{Surface Brightness profile of Diffuse X-ray Emission}  
We derived  azimuthally averaged, background subtracted, 0.3-2.5~keV radial
surface brightness profile by extracting X-ray photons from 35 concentric
circular annuli extending upto 360$\arcsec$ centered on
the CTD~86 using the task {\it funtool} available in ds9.  To avoid any
contamination due to the central source, we excluded the central
15$\arcsec$ region during the analysis.  The width of the annuli, particularly
in the outer parts, were adjusted so as to achieve roughly the same
signal-to-noise ratio.  The radial surface brightness profile thus derived is
shown in Figure~\ref{sur}. From this figure it is apparent that the extended
emission is detectable upto a radius of  $\sim$250\arcsec($\sim$ 186~ kpc).
This profile was fitted with a single  1-d $\beta$ model
\citep{1962AJ.....67..471K} that resulted in the best fit core radius
$r_{c}$=46.12$\pm$4.46~kpc and the slope parameter $\beta$ = 0.61$\pm$0.03.  The
errors given here are at $68\%$ confidence level.

\subsection{X-ray Spectral Analysis}
\label{spec}

\begin{figure}
\centering
\includegraphics[width=120mm]{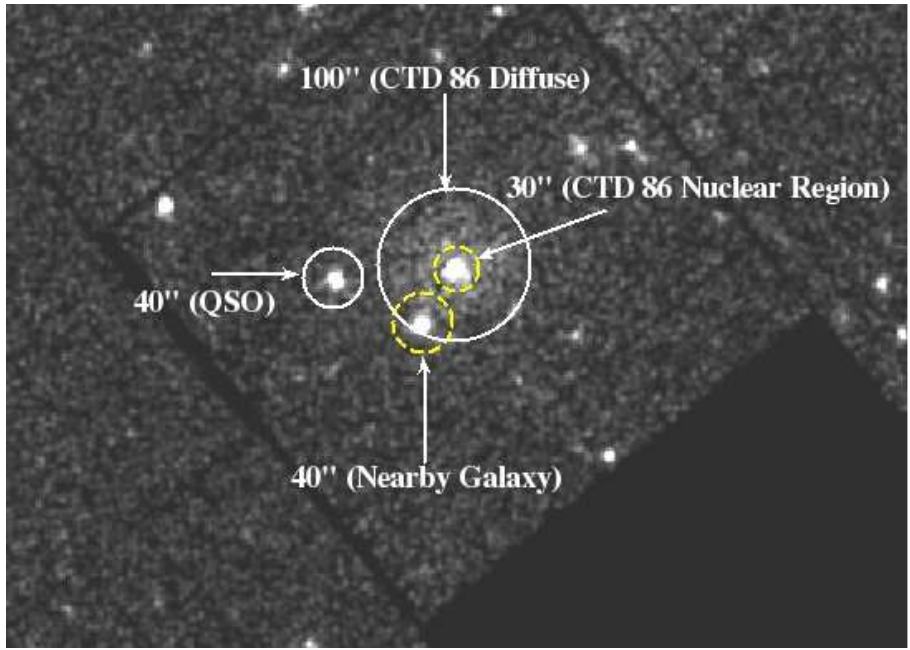}
\caption{Background subtracted and  3\arcsec Gaussian smoothed MOS1 image of the region containing CTD~86.  Superimposed over the image are the regions used for spectral
extraction as described in $\S$~\ref{diff}.  A 100\arcsec circular region (white) centered on CTD~86 was used to extract spectrum for the diffuse emission,
excluding the regions shown as dotted circles in yellow. For QSO (SDSS
J142452.11+263715.1) and nearby galaxy (SDSS J142443.78+263616.2) 40\arcsec
regions as indicated in the figure were used for spectral extraction.}

\label{extraction}
\end{figure}

Exposure-corrected, 3$\arcsec$ Gaussian smoothed image of \xmm{} is shown in
Figure~\ref{extraction}. We extracted EPIC-pn, MOS1 and MOS2 spectra of the hot
diffuse gas around CTD~86 using a circular region with radius $100\arcsec$ but
excluding two circular regions one of radius $30\arcsec$ centered on CTD~86 and
another of radius of $40\arcsec$ centered on the nearby source
SDSS~J142443.78+263616.2 (figure~\ref{extraction}).  We also extracted the EPIC
spectra of the nuclear source using a circular region with a radius of
$30\arcsec$ centered at the position of CTD~86.  The background spectra were
extracted from the source-free regions.  Spectra for both the bright X-ray
sources SDSS~J142443.78+263616.2 and SDSS~J142452.11+263715.1 were also
extracted using circular regions of radii $40\arcsec$ centered on their peak
positions. Ancillary response files (ARF) and
redistribution matrix files (RMF) were generated for each of the extracted
spectrum using the SAS tasks {\tt rmfgen} and {\tt arfgen}, respectively.  The
resulting spectra were binned to a minimum of 30 counts per bin for CTD~86 and
SDSS 142443.78+263616.2, and 20 counts per bin for SDSS J142452.11+263715.1.  
We used the X-ray spectral analysis package
XSPEC v12.6.0 (\citealt{1996ASPC..101...17A}). 

We analyzed the EPIC-pn, MOS1 and MOS2 spectral data jointly for each
of the three sources.  Several models, discussed in $\S$~\ref{diff}, were
used along with a constant component in order to account for possible
differences in the relative normalizations of the three instruments.
The constant component was fixed at unity (1) for the EPIC-pn data and varied
for the MOS1 and MOS2 data. The best fit parameters were
derived using $\chi^2$ minimisation technique and the errors were derived
for the 90\% confidence level.  The details of the models fitted  and the
resulting parameters along with their errors are given below.

\subsubsection{The diffuse emission}
\label{diff}
Spectral data from the region dominated by diffuse emission around CTD~86, were
analyzed using a single component model {\tt apec},
appropriate for a thermal plasma, modified by the Galactic absorption
($N_{H}^{Gal} =1.61\times10^{20}{\rm~cm^{-2}}$).  The best fit
resulted in minimum $\chi^2=378.06$ for $278$ degrees of freedom (dof).
 The residuals of the fit showed presence of weak excess
emission at higher energies and a residual instrumental line at $1.79\kev$ due
to Si~K$\alpha$. Therefore, we added a power-law to account for the
contribution from the large point spread function of the nuclear emission and a
Gaussian component to account for the  $1.79\kev$  line.   It may also have some
contribution from the incorrect subtraction of the background due to soft
protons \citep{2008A&A...478..615S, 2011ApJ...743...78L}. 
The {\tt wabs(apec+powerlw+Gauss)} model improved the fit to $\chi^2/dof
=283.59/274$.  The spectral parameters for the powerlaw component were poorly
constrained, hence we decided to fix the photon index and performed spectral
fitting by keeping it fix at 1.5, the fit results in to $\chi^2/dof
=303.11/275$. This resulted in the best fit temperature of the hot gas
$kT= 0.79\pm0.02\kev$ and elemental abundance $0.29\pm0.05$ relative to
the solar values. 
We also performed spectral analysis of the data resulting
from the Obs.II and found that the data are well described by a similar model.
The spectral analysis resulted in to the $0.3-10\kev$  X-ray
flux of diffuse component to be 
$15.1\pm0.7\times10^{-13}{\rm~erg~s^{-1}~cm^{-2}}$ and
$13.0\pm0.6\times10^{-13}{\rm~erg~s^{-1}~cm^{-2}}$ for Obs.I and II,
respectively.  The best fit values for the elemental abundances for Obs.I
and Obs.II are found to be $0.29\pm 0.05$ $Z_{\rm\odot}$ and $0.31\pm 0.04$
$Z_{\rm\odot}$, respectively.  The $0.3-10\kev$ X-ray luminosity of the diffuse
emission from CTD~86  is thus found to be $4.6\pm0.2\times10^{42}
{\rm~erg~s^{-1}}$ and $4.0\pm0.2\times10^{42} {\rm~erg~s^{-1}}$
for Obs.I and II, respectively. The best fit parameters are listed in
Table~\ref{xray_spec_par_ctd86} and the spectral data, the
best-fit model and the deviations compared to the best fit model are
shown in Figure~\ref{xray_spec_plot_ctd86}.  The allowed values of  temperature,
abundance and $N_H^{intr}$ for Obs.I and Obs.II along with their confidence
  contours at the 68.3\%, 90\%, and 99\% confidence levels are shown in
Figure~\ref{cont}.  The  electron density of the thermal plasma
can
be calculated from the normalization of {\tt apec} model,

\begin{equation}
   n_{apec} = \frac{10^{-14}}{4\pi[D_A (1+z)]^2}\int{n_e n_H dV},
\end{equation}

where $D_A$ is the angular diameter distance to the source, $n_e$ and
$n_H$ are the electron and proton densities, respectively.  For
CTD~86, using the best-fit temperature $0.79\pm0.02\kev$ and assuming the
uniform gas density within the central 100\arcsec $\sim$ 74.4 kpc) and $n_e
\approx n_H$, we find the average electron density $n_e \approx
3\times10^{-3}{\rm~cm^{-3}}$ and the total diffuse gas  mass ${\rm M_{gas}}=
1.83\times10^{11}~{\rm\Msun}$.

\subsubsection{The nuclear X-ray source}
\label{cent}
\begin{figure*}
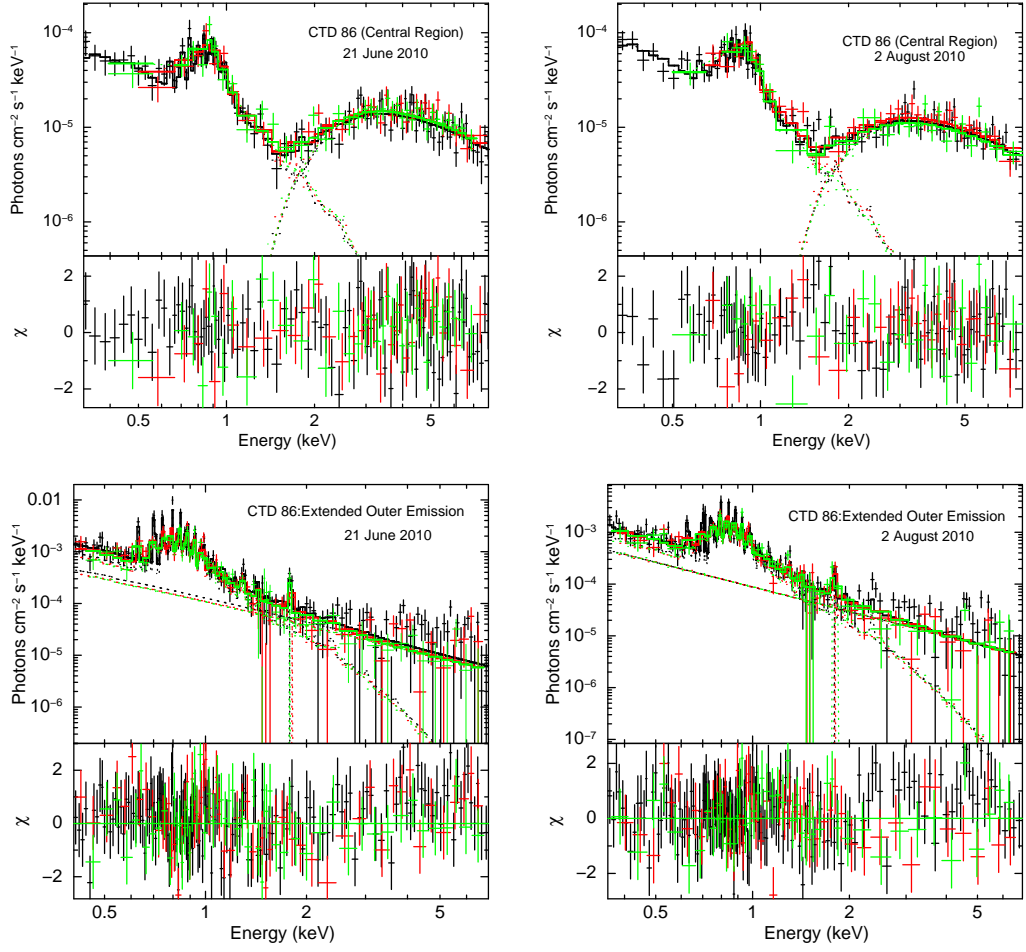

  \vbox {
    \includegraphics[width=70mm]{ctd86_fig6a.ps}
    \includegraphics[width=70mm]{ctd86_fig6b.ps}
    \includegraphics[width=70mm]{ctd86_fig6c.ps}
    \includegraphics[width=70mm]{ctd86_fig6d.ps}
  }
  \caption{X-ray spectral data, best-fit models and the deviations of the data
from the best fit models for CTD~86 from two observations: Obs.I (left)  \&
Obs.II (right).}
  \label{xray_spec_plot_ctd86}
\end{figure*}

\begin{figure*}
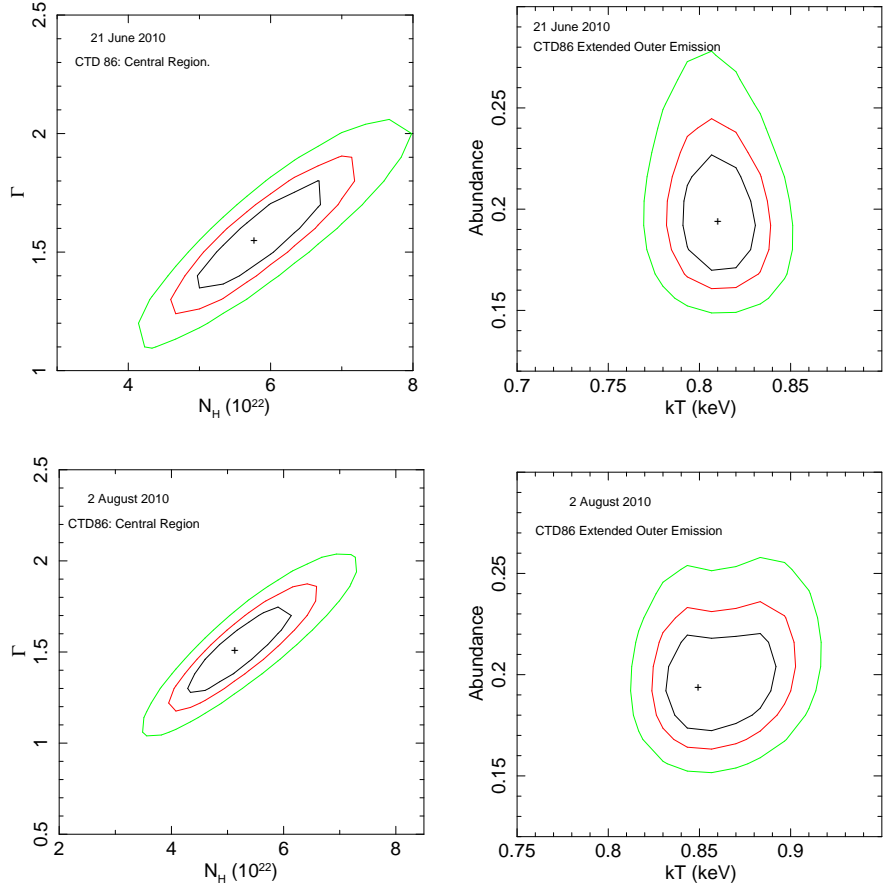

  \scriptsize \vbox {
    \includegraphics[width=60mm]{ctd86_fig7a.ps}
    \includegraphics[width=60mm]{ctd86_fig7b.ps}
    \includegraphics[width=60mm]{ctd86_fig7c.ps}
    \includegraphics[width=60mm]{ctd86_fig7d.ps}
  }
  \caption{{\it Top left:}
    $\chi^2$ contours of the photon index, $N_H^{intr}$ measurements
    for the central 30\arcsec regions of CTD 86, {\it top right
      :} $\chi^2$ contours of the temperature and abundance
    measurements for the 100\arcsec region excluding central 30\arcsec
    from the spectral analysis of Obs.I. {\it
      bottom left :} $\chi^2$ contours of the photon index,
    $N_H^{intr}$ measurements for the central 30\arcsec regions of CTD
    86. {\it bottom right:} $\chi^2$ contours of the temperature
    and abundance measurements for the 100\arcsec region excluding
    central 30\arcsec from the spectral analysis of data for Obs.II . 
    The confidence levels for the innermost, middle,
    and outermost contours are at 68.3\%, 90\%, and 99\%,
    respectively.}
  \label{cont}
\end{figure*}

X-ray emission from the central part of CTD~86 contains contribution from
the nuclear source (AGN) and the hot gas surrounding it.  Consequently,
we performed spectral analysis of X-ray photons from the central region of
CTD~86 using a two component model consisting of an {\tt apec}, appropriate for
a thermal plasma, plus a {\tt powerlaw} for the central AGN modified by the
Galactic absorption ($N_{H}^{Gal} = 1.6\times10^{20}{\rm~cm^{-2}}$).  The fit
resulted in $\chi^2=345.7$ for $167$
dof for Obs.I. Examination of the residuals showed a deficit of emission in the
$1.5-3\kev$ band, suggestive of intrinsic absorption of the power-law component.
  Therefore, we included an additional absorption component {\tt zwabs}, that
improved the fit to $\chi^2/dof$ = $159.8/164$.  We also performed spectral
analysis of the data resulting from the Obs.II and found that a similar model
was required to fit the data. The best-fit parameters of the fits are listed in
Table~\ref{xray_spec_par_ctd86} and the plots are shown in
Figure~\ref{xray_spec_plot_ctd86}.  The allowed range of values are shown in
Figure~\ref{cont} as enclosed by confidence contours at the 68.3\%, 90\%, and
99\% confidence levels for temperature, abundance and $N_H^{intr}$ for Obs.I and
Obs.II. We note that 
 
The model {\tt wabs(apec+zwabs$\times$powerlaw)} provided the
intrinsic absorption column $N_H = 5.9_{-1.0}^{+0.8}\times 10^{22} 
{\rm~cm^{-2}}$ and $5.4_{-1.0}^{+1.3}\times 10^{22} {\rm~cm^{-2}}$ for 
Obs.I and II, whereas the power law indices ($\Gamma$)
 derived for Obs.I and II are $1.58_{-0.25}^{+0.27}$, $1.54_{-0.27}^{+0.29}$, 
respectively.  The best fit values for the elemental abundance for the
hot gas component for both the observations are found to be $0.15\pm
0.05$ and $0.14\pm 0.07$ relative to the solar values.  The temperature
of the diffuse emission is $0.89\pm0.05\kev$, and the corresponding diffuse
X-ray flux in the $0.3-10\kev$ band is 
$0.66\pm0.10\times10^{-13}{\rm~erg~s^{-1}~cm^{-2}}$ and
$0.71\pm0.12\times10^{-13}{\rm~erg~s^{-1}~cm^{-2}}$ for Obs.I and II,
respectively.  This results in to the $0.3-10\kev$  X-ray
luminosities to be $2.1\pm0.2\times10^{42} {\rm~erg~s^{-1}}$ and
$1.8\pm0.3\times10^{42} {\rm~erg~s^{-1}}$ for Obs.I and II, respectively.
Using the best-fit temperature $0.89\pm0.05\kev$, and assuming the uniform gas
density within the central region of radius 30\arcsec ($\sim$ 0-22 kpc), we got
$n_e \approx 4.4\times10^{-3}{\rm~cm^{-3}}$  and a total gas mass
${\rm M_{gas}}=5.9\times10^{9}~{\rm\Msun}$.

\begin{table*}
\scriptsize
  \caption{Results of X-ray spectral analysis of CTD~86 for Obs.I $\&$ Obs.II.}
  \begin{tabular}{lcccccccc} \hline
    Region  &~~~~~~~~~~~~~~~~~~~~`` Central($r\le30\arcsec$)    &
&~~~~~~~~~~~~~~~~~~~~~~~~Diffuse ($30\arcsec\le r \le 100\arcsec$)  &
            \\\hline
    Model  & \multicolumn{2}{c}{wabs(apec+zwabs$\times$ZPL)}&
\multicolumn{2}{c}{wabs(apec+PL+Gauss)}    \\
    & Obs.I  & Obs.II                     & Obs.I & Obs.II
\\ 
    & (21 June 2010) & (2 Aug. 2010)    & (21 June 2010)& (2 Aug. 2010)  \\
\hline

    $N_H^{Gal}$($10^{20}{\rm~cm^{-2}})$        &1.6(fixed)
   &1.6(fixed)	              &1.6(fixed)              &1.6(fixed)    	\\ \\
    $kT_{apec}$($\kev$)                        &$0.89\pm0.05$
   & $0.89\pm0.07$           &$0.79\pm0.02$           &$0.81\pm0.01$   \\
    Abundance ($\times$ solar) 		       &$0.15\pm0.05$
   & $0.14\pm0.07$           &$0.29\pm0.05$           &$0.31\pm0.04$    \\
    $f_X^{apec}$($0.3-10\kev$)$^a$             &$0.66\pm0.10$
   & $0.71\pm0.12$           &$15.1\pm0.7$            &$13.0\pm0.6$    \\ \\
    $N_H^{intr}$($10^{22}{\rm~cm^{-2}}$)       &$5.9_{-1.0}^{+0.8}$
   &$5.4_{-1.0}^{+1.3}$       &--&--\\	         
    $\Gamma$                                   &$1.58_{-0.25}^{+0.27}$
   &$1.54_{-0.27}^{+0.29}$    &1.58                    & 1.58    \\
    Si~K$\alpha$ $G_{line}$($\kev$)            &  --
   & --                       &$1.77\pm0.03$           &$1.77\pm0.04 $     \\
    $\sigma(gauss)($\kev$)$                    & --
   &  --                      &$4.01\times10^{-4}$     &$4.07\times10^{-4}$  \\
    {\it f} {Si~K$\alpha$}$^b$                 & --
   &  --                      &$3.82\pm4.62\times10^{-6}$
&$4.61\pm3.39\times10^{-6}$                   \\
    $f_X^{PL}$($2-10\kev$)$^a$                 & $8.3_{-1.6}^{+0.9}$
   & $6.7_{-0.4}^{+0.5}$      &--                      &--               \\ \\ 
    $\chi^2/dof$	                       &159.8/164
   &134.9/140                 &303.11/275               &249.38/253       \\ 
    \hline
    \multicolumn{2}{l}{$^a$ In units of $10^{-13}{\rm~erg~cm^{-2}~s^{-1}}$.}\\
        \multicolumn{2}{l}{$^b$ Line flux in photons
${\rm~keV^{-1}~cm^{-2}~s^{-1}}$.}
   \end{tabular} 
  \label{xray_spec_par_ctd86}       
\end{table*}

\begin{figure*}
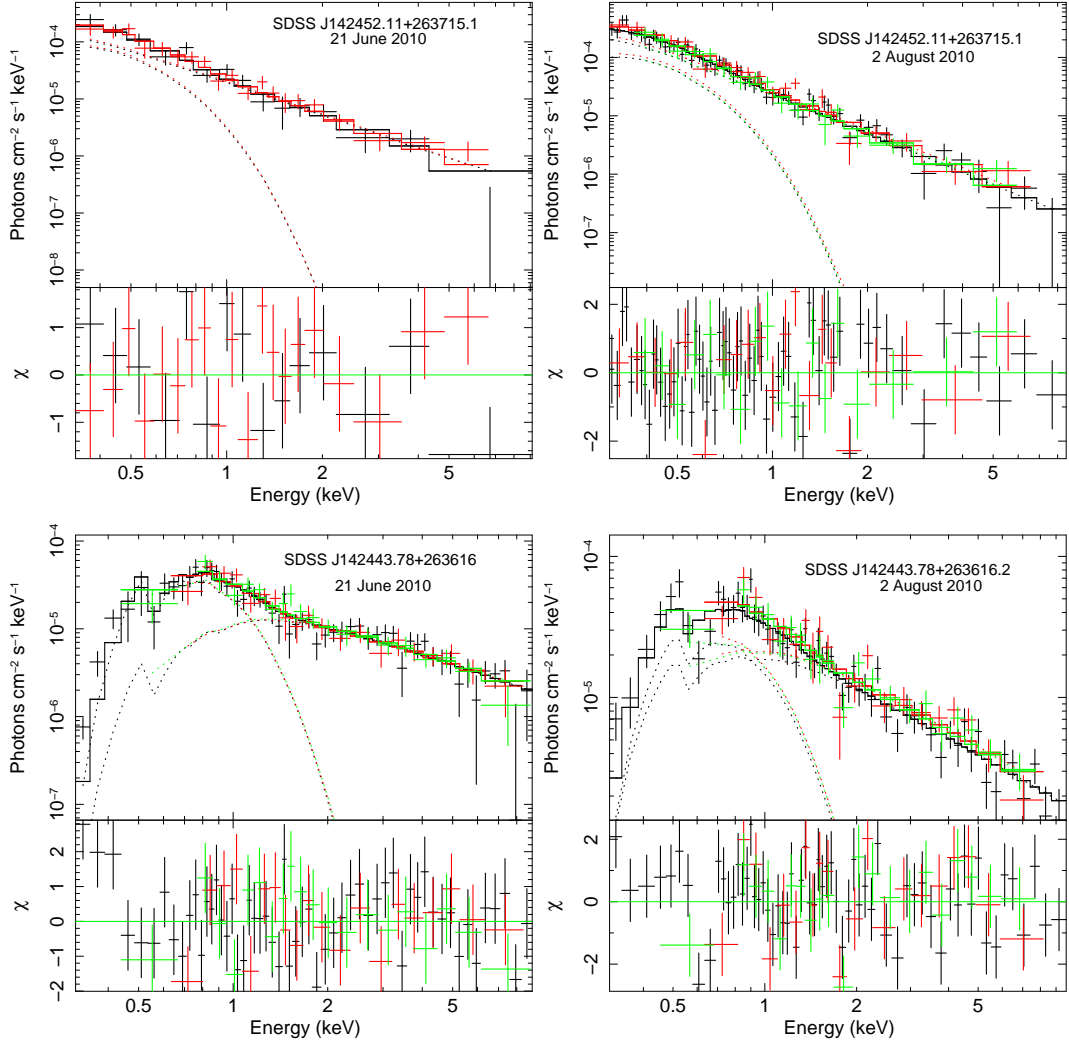

  \vbox {
    \includegraphics[width=70mm]{ctd86_fig8a.ps}
    \includegraphics[width=70mm]{ctd86_fig8b.ps}
    \includegraphics[width=70mm]{ctd86_fig8c.ps}
    \includegraphics[width=70mm]{ctd86_fig8d.ps}
  }
  \caption{X-ray spectral data, best-fit models and the deviations of
    the data from the best fit models for SDSS~J142452.11+263715.1(upper panel)
and SDSS~J142443.78+263616.2 (lower panel) from two observations: Obs.I (left)
\& Obs.II (right).}
  \label{xray_spec_plot_point_sources}
\end{figure*}

\begin{table*}
 \scriptsize
  \caption{Best-fit spectral model parameters of the two nearby point sources
for Obs.I $\&$ Obs.II}
  \begin{tabular}{lcccccccc} \hline
    Parameter &  \multicolumn{2}{c}{SDSS~J142452.11+263715.1} &
\multicolumn{2}{c}{SDSS~J142443.78+263616.2} \\\hline
    Model  & \multicolumn{2}{c}{wabs(ZBB+ZPL)}             &
\multicolumn{2}{c}{wabs*wabs(BB+PL)}  \\
    &                 Obs.I & Obs.II       	     &Obs.I  & Obs.II   \\ 
    &  (21 June 2010)& (2 Aug. 2010)   &(21 June 2010) & (2 Aug. 2010) \\ \hline

    $N_H^{Gal}$($10^{20}{\rm~cm^{-2}})$          &1.6(fixed)   	&1.6(fixed)
&1.6(fixed) &1.6(fixed)	\\ \\
  
    $kT_{BB}$ ($\kev$) 	                        &$0.15_{-0.07}^{+0.04}$
&$0.14_{-0.04}^{+0.02}$ &$0.13_{-0.02}^{+0.04}$ & $0.17_{-0.04}^{+0.07}$ \\
    $f_X^{BB}$($0.3-10\kev$)$^a$               & $0.23_{-0.19}^{+0.17}$ &
$0.24_{-0.15}^{+0.13}$ & $2.3_{-1.7}^{+7.4}$ & $0.6_{-0.3}^{+1.6}$ \\ \\
    $N_H^{intr}$($10^{22}{\rm~cm^{-2}}$)       &--&--&$0.36_{-0.20}^{+0.22}$
&$0.16_{-0.09}^{+0.34}$\\	         
    $\Gamma$                                    & $1.95_{-0.35}^{+0.34}$
&$2.22_{-0.23}^{+0.23}$& $1.20_{-0.20}^{+0.21}$ & $1.24_{-0.18}^{+0.18}$\\

    $f_X^{PL}$($2-10\kev$)$^a$                 &   $0.55_{-0.16}^{+0.19}$ &
$0.43_{-0.09}^{+0.10}$ & $2.5\pm0.3$ & $2.5\pm0.2$\\ \\ 
    $\chi^2/dof$	                       & 30/30& 105/98 &83/89 &125/98
\\ 
    \hline
    \multicolumn{2}{l}{$^a$ In units of $10^{-13}{\rm~erg~cm^{-2} s^{-1}}$.}
  \end{tabular} 
  \label{xray_spec_par_point_sources}       
\end{table*}

\subsection{Bright X-ray Sources}
Two bright X-ray sources: SDSS J142452.11+263715.1 and SDSS J142443.78+263616.2
were evident near CTD~86 at a projected distance 
of $119{\rm~kpc}$ and $64{\rm~kpc}$, respectively from the centre of CTD~86
(see Fig.~\ref{xray_color}).  In the SDSS
DR7 and DR8 data, SDSS~J142452.11+263715 has been identified as a star
and SDSS~J142443.78+263616.2 as a galaxy.  Based on optical
identification and analysis of the SDSS optical spectrum, we find that
the star-like object SDSS J142452.11+263715 is in fact a broad emission line
quasar with $z=0.3761$ (Pandge et al. 2013 in preparation).  The nature of the
other source, SDSS~J142443.78+263616.2, is not clear as its optical spectrum is
not available.  This object is appears as an extended object in the SDSS image
and is likely a galaxy harboring an AGN.

To further examine the nature of these sources, we performed an X-ray spectral
analysis of each source.  For this, we extracted source spectra from
the filtered EPIC-pn and MOS event files using circular regions of
40\arcsec~radii centered at the source positions.  The corresponding background
spectra were extracted from the source-free regions.  The source SDSS
J142452.11+263715.1 falls in the CCD gap in the EPIC-pn data of Obs.I, hence the
EPIC-pn spectrum was not useful.  The rest of the five spectra from Obs.I and II
were grouped to get minimum counts of 20 per energy bin.  As before, we
performed joint spectral analysis of available spectral data from different
instruments for each observation.
  
We found that the spectra of SDSS J142452.11+263715.1 for each
observation are well constrained by a black body plus power-law model
modified by the Galactic absorption ($N_H^{Gal}=1.6\times10^{20}{\rm~cm^{-2}}$)
and the best-fit parameters are listed in
Table~\ref{xray_spec_par_point_sources}.  The X-ray spectrum of SDSS
J142452.11+263715.1 consisting of soft X-ray excess, described as a black body
with $kT\sim150\ev$ and a powerlaw of $\Gamma \sim 2$, is typical of the spectra
of radio-quiet quasars e.g., PG quasars.  The total X-ray luminosity of SDSS
J142452.11+263715.1 in the $0.3-10\kev$ band is $6.6\pm0.3\times10^{43}
{\rm~erg~s^{-1}}$ and  $6.8\pm0.25\times10^{43} {\rm~erg~s^{-1}}$
for Obs.I and II, respectively.  The best fitted spectra of SDSS
J142452.11+263715.1 for both the observations are
shown in Figure~\ref{xray_spec_plot_point_sources} (upper panels).

The second object SDSS J142443.78+263616.2 has no redshift
information.  This source shows soft X-ray emission.  Initially we used
a model {\tt wabs$\times$wabs({apec+PL})}, where thermal plasma component {\tt
apec} is for the soft excess and powerlaw for the hard continuum. We varied the
red shift parameter in {\tt apec}.  The fit resulted in $\chi^2/dof = 132/88$
and the best-fit redshift $z=1.50\pm0.05$, which may not be correct.  Next, we
fitted the {\tt wabs$\times$wabs({BB+PL})} model, which resulted in $\chi^2/dof
=83.70/89$, $125/98$ for Obs.I and II, respectively. The best-fit
parameters are $kT_{BB} \sim 150\ev$ and $\Gamma \sim 1.2$ for both
the observations.  The best fitted spectra of SDSS J142443.78+263616.2
for Obs.I and II are shown in Figure~\ref{xray_spec_plot_point_sources} (lower
panels) and the spectral fitting parameters are listed in
Table~\ref{xray_spec_par_point_sources}.

\section{Discussion}

CTD~86 shows extended X-ray emission with a linear size $\sim$
250\arcsec (186~kpc) in the \xmm{} observations, confirming the
previous observation by \citet{1999MNRAS.310...30C} using {\it ROSAT} data.  
Diffuse X-ray emission detected with the \xmm{} is well described by thermal
emission from hot gas with a temperature $kT=0.79\pm0.02\kev$, electron density
$n_e\sim 3\times10^{-3}{\rm~cm^{-3}}$ and $0.3-10\kev$ X-ray luminosity
$L_X=4.8\pm0.2\times10^{42}{\rm~erg~s^{-1}}$, implying that this system is very
likely to be part of a group of galaxies or a poor cluster.  
The spectral analysis of the diffuse gas component yielded the metal content of
the CTD~86 to be $0.29\pm 0.05$ and $0.31\pm 0.04$, respectively, these values
are in agreement with those reported earlier by \cite{2010ApJ...710.1205H} using
10\,ks {\it Chandra} data.  Such subsolar metallicities are quite common, as
reported by \cite{2005MNRAS.357..279C} and \cite{2000MNRAS.315..356H}, for
intragroup medium around several FR-I radio galaxies or gas in groups of
galaxies based on measurments with {\it ROSAT} PSPC data.

For both the observations, X-ray emission from the low luminosity AGN in
CTD~86 is well described by a power-law ($\Gamma\sim1.55$) modified by
intrinsic absorption with $N_H\sim5.5\times10^{22}{\rm~cm^{-2}}$.  
In a similar analysis, \citep{2007PThPS.169..278M} found high absorption
column ($N_H \sim 1.5\times10^{23}{\rm~cm^{-2}}$) for the FR-I radio galaxy
Centaurus A.   Thus, the intrinsic absorption column of
CTD~86 is smaller by a factor of $\sim2.5$ than that in  Centaurus A.  For a
sample of FR-I galaxies \citep{2004ApJ...617..915D} observed a range of photon
indices, $\Gamma = 1.1-2.6$, and intrinsic absorption column densities,
$N_H=10^{20}-10^{21}{\rm~cm^{-2}}$.  The power-law photon index derived for
CTD~86 is found to lie in the range for FR-I type galaxies. The
$2-10\kev$ luminosity of the nucleus is found to be
$L_X =  2\times10^{42}{\rm~erg~s^{-1}}$, making CTD~86 a low luminosity AGN.
Based on the low column densities observed for a sample of FR~I galaxies,
\cite{2004ApJ...617..915D} concluded that most FR~I galaxies lack a standard
torus. Since ISM of elliptical galaxies are unlikely to provide absorbing
columns more than $10^{21}{\rm~cm^{-2}}$, therefore, CTD~86 provides an
intriguing possibility of the presence of a torus around the FR~I galaxy.

For the case of CTD~86, we investigated the existence of Fe
${\rm K}_\alpha$ line in the nuclear spectrum and found that
the line is not detected at  $3\sigma$ level.
Therefore, we calculated the 90\% upper limit on the equivalent width of a narrow, neutral Fe ${\rm K}_\alpha$ line  to be $66$~eV. This upper limit is a factor of $\sim 1.5$ lower than the equivalent width of  $\sim$100~eV measured for Cen~A  \citep{2004ApJ...612..786E}. The weaker iron line in CTD~86 could be due to a lower iron abundance or lesser intrinsic absorption column compared to that in Cen~A. 
However, the upper limit on the iron line equivalent width of CTD~86
is comparable to that observed from an intermediate Seyfert galaxy
MCG-5-23-16 (EW$\sim 40$~eV) with intrinsic $N_H\sim 1.6\times
10^{22}{\rm~cm^{-2}}$ \citep{2003ApJ...592...52D} or Seyfert 2/LINER
NGC~4258 (EW$\sim65$~eV) with cold $N_H \sim 10^{23} {\rm~cm^{-2}}$
\citep{2000ApJ...540..143R}.  Thus, the upper limit on the equivalent
width of the Fe ${\rm K}_\alpha$ line from CTD~86 is similar to that
typically found for intermediate type Seyfert galaxies with comparable
intrinsic absorption column.

VLBI imaging of CTD~86 has revealed symmetric radio jets on a kpc scale and a
possible one-sided jet on parsec scale \citep{2005ApJ...618..635G}.  Assuming
the parsec scale structure to be real, \cite{2005ApJ...618..635G} estimated an
inclination angle i = 45-50\degree and $\beta$=0.95.  X-ray
absorption column of $N_H \sim 5.9\times10^{22}{\rm~cm^{-2}}$ in
CTD~86 is similar to the absorption columns usually observed for
intermediate type Seyfert galaxies (e.g., Seyfert 1.5).  Thus, the
absorption column and the tentative inclination angle both are
consistent with the presence of a torus where the line of sight to the
nucleus passes through the outer region of the torus.

It is interesting to investigate the disk-corona geometry in CTD~86 and to check
its similarity with that in type 1 radio-quiet AGN like Seyfert 1 galaxies.  We
 measured the stellar velocity dispersion $\sigma = 182\pm8\kms$of CTD~86
(Pandge et. al 2013 in preparation) using the Penalized Pixel Fitting method
\citep{2004PASP..116..138C} from the SDSS spectrum  and calculated the black
hole mass, $M_{BH} = (8.8\pm2.4)\times10^{7}{\rm~M_{\odot}}$ using the
$M_{BH}-\sigma$ relation.  We estimated the bolometric luminosity from
the $2-10\kev$ luminosity assuming a bolometric correction of 20
\citep[e.g.,][]{2007MNRAS.381.1235V}.  This resulted in $\dot{m} =
L_{bol}/L_{Edd} \sim 4\times10^{-3}$ quite low compared to that found for
typical Seyfert and quasar like objects.  One way to find the relative
contributions of disk and corona is to compare the ratio of luminosities of
[O~III]$\lambda5007$ line and the unabsorbed $2-10\kev$ luminosity.  The
strength of the [O~III] line would depend on the ionizing flux i.e., emission
from the accretion disk, while the power-law X-ray emission arises from the hot
corona. \cite{2004ApJ...617..915D} found  $L_X/L_{[O~III]}$ $\sim$ $27.8\pm 9.6$
for Seyfert 1 galaxies and $3.7\pm1.1$ for Compton thick Seyfert 2 galaxies,
where as the lack of X-ray emission from Seyfert 2 galaxies was attributed to
the absorption by Compton-thick torus.  We find a very large, $L_X/L_{[O~III]}=
354$, ratio for CTD~86, which either suggests a beamed X-ray
emission arising from the jet or a lack of ionizing luminosity from the
accretion disk. The first possibility, i.e. the jet origin, is unlikely as the
X-ray emission is absorbed by a high column density ($N_H \sim
10^{22}{\rm~cm^{-2}}$). The second scenario is possible only if the accretion
disk is truncated so that inner disk does not exist and there is no ionizing
radiation from the hottest part of the disk.  The inner region below the
truncation radius may be filled with a hot flow that gives rise to the observed
X-ray emission.  The inner hot flow may also be responsible for the radio jets.

\begin{figure}
  \includegraphics[width=85mm]{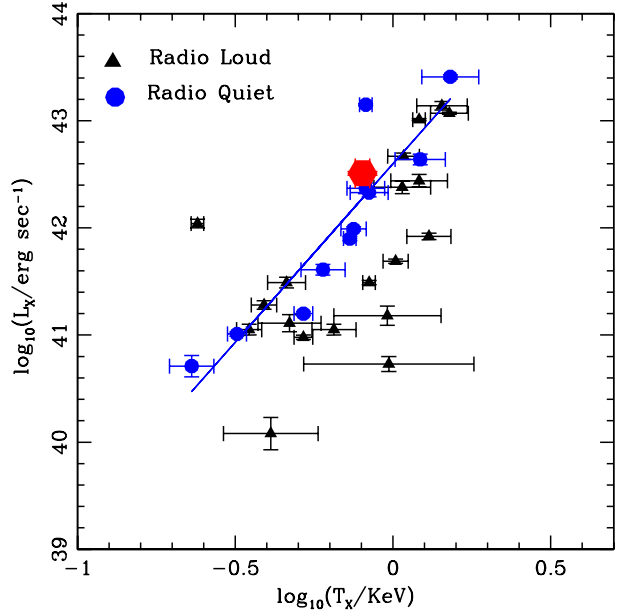}
  \centering
  \caption {$L_X$ plotted against the temperature of the hot gas for a
    sample of galaxies from \citet{2005MNRAS.357..279C}. The filled
    triangles represent radio-loud galaxies and the filled circles
    represent radio-quiet galaxies. CTD~86 is shown as the filled
    hexagon.}
  \label{Lx_R}
\end{figure}

\begin{figure}
  \centering
  \includegraphics[width=8cm]{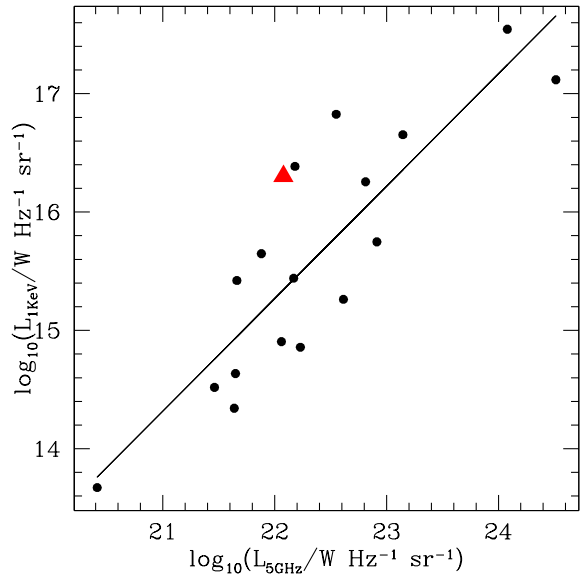}
  \caption{Relationship between the flux densities at $1\kev$ and  $5{\rm~GHz}$
 for FR~I galaxies taken from \citet{2006ApJ...642...96E}. CTD~86 is shown by
the triangle.}
  \label{radio_xray_core}
\end{figure}

CTD~86 is a low luminosity radio galaxy with classical double-lobed
radio emission.  Figure~\ref{xray_opt_radio} revealed that the lobes
extend up to $\sim$ 65 kpc from the optical center of the galaxy and the
entire radio source is well contained in the hot X-ray emitting
gas.  The distribution of hot gas in groups and clusters is  affected by the
presence of radio jets and lobes. Several example of cavities in the
distribution of X-ray gas, coinciding with the locations of radio lobes have
been observed with \chandra{} and \xmm{}
\citep[e.g,][]{2007ARA&A..45..117M,2009ApJ...705..624D,
2010ApJ...712..883D,2012MNRAS.421..808P,2013Ap&SS.345..183P}. 
Though, radio lobes from CTD~86 are found to be embedded in the hot gas,
however, there is no clear evidence for the distortions in the radio
structures as well as in the hot gas in the form of X-ray
cavities.  \citet{2005MNRAS.357..279C} studied radio source heating in groups
and found that the gas around the radio-loud AGN is more likely to be hotter at
a given X-ray luminosity compared to the gas in radio-quiet groups.  To confirm
this, we compared the X-ray luminosity and temperature
of extended gas around CTD~86 with that of other radio loud and radio
quiet galaxies of similar type.  Figure~\ref{Lx_R} gives a correlation between
the $\log{L_{X}}$ versus $\log{T_{X}}$ for a sample of galaxies taken
from \citet{2005MNRAS.357..279C}.  The filled triangles (black) represent data
points for radio loud sample and filled circle (blue) represent sample of radio
quiet galaxies.  Most of the radio loud systems are found  to lie below the best
fit for radio quiet groups of galaxies.  In the same plot, we have shown CTD~86
with a red hexagon.  The temperature and luminosity values of the hot gas in
CTD~86 environment are consistent with the $L_{X} -T$ relation for the
radio-quiet groups or galaxies.  Thus, the $L_{X} - T$ relation, the absence of
X-ray cavities, symmetric radio structures of CTD~86 all suggest that heating by
the radio source in this system is not significant, probably due to the low
power of the radio galaxy.
Figure~\ref{radio_xray_core} shows a correlation
between luminosity
densities at $5{\rm~GHz}$ and $1\kev$ for a sample of FR~I galaxies taken from
\citet{2006ApJ...642...96E}.  In this figure, we also show the position of
CTD~86 which is not close to the correlation line for the FR-I galaxies. The
X-ray luminosity of CTD~86 is higher by an order of magnitude compared to other
sources of comparable radio emission.  Thus, based on this correlation, it is
unlikely that the X-ray emission from CTD~86 galaxies arise from the jets
alone.  X-ray emission from CTD~86 is also intrinsically absorbed by a
large column within the host galaxy which is not likely if the entire
X-ray emission were to arise from the jet.  We checked for the
presence of an intrinsically unabsorbed X-ray power-law component that
could arise from the jet. The contribution of such a power-law is at the
most $35\%$ of the total X-ray emission in the $2-10\kev$ band,
therefore, most of the X-ray emission above $2\kev$ might be arising
from the hot corona.

\section{Conclusions}
We have presented X-ray images and spectra of the
CTD~86 and its environment using the publicly available \xmm{} observations.
The main results of the
study are:

\begin{enumerate}
\item An extended ($\sim186{\rm~kpc}$) X-ray
emission from CTD~86 with spectral properties $kT\sim 0.79\kev$, $n_e\sim
10^{-3}{\rm~cm^{-3}}$, ${\rm M_{gas}}= 1.83\times10^{11}~{\rm\Msun}$ and
Z=$0.29\pm 0.05$ $Z_{\rm\odot}$, is seen, and is similar to that found in other
group galaxies.

\item No signs of interaction between the radio lobes and the hot
gaseous environment of CTD~86 are evident in the form of X-ray cavities.

\item  Mild nuclear activity with $2-10\kev$ X-ray luminosity of
 $\sim$ $2.0\times10^{42}{\rm~erg~s^{-1}}$ is evident in CTD~86.

\item The Nuclear X-ray spectrum of  CTD~86 absorbed by a larger column ($N_H
= 5.9\times10^{22}{\rm~cm^{-2}}$) and may be due to the presence of a torus
around CTD~86, similar to that seen in other type 2 AGN.

\item The large $L_X/L_{[O~III]}$ ratio and very small relative  accretion rate
of CTD~86 suggest that the accretion disk of this low luminosity AGN is likely
truncated at some large inner radius, where the inner regions are filled with a
hot flow.

\item Two bright X-ray sources, SDSS~J142452.11+263715.1 and
SDSS~J142443.78+263616.2, near CTD~86 are reported here.  
SDSS~J142452.11+263715.1 has been identified as a star like object,
probably harbors a type 1 AGN with $z=0.3761$ and has unabsorbed $0.3-10\kev$
X-ray luminosity $L_X\sim 8 \times 10^{43}{\rm~erg~s^{-1}}$, while SDSS
J142443.78+263616.2 is probably a galaxy hosting an AGN.
\end{enumerate}

\section*{Acknowledgments}
MBP gratefully acknowledge support by the DST, New Delhi under the INSPIRE
fellowship program (sanction No. IF10179).  MBP and MKP acknowledge the usage
of high performance computing facilities procured under the DST-FIST scheme
(SR/FST/PSI-145). This work is based on observations obtained with XMM-Newton,
ESA science mission with instruments and contributions directly funded by ESA
Member States and NASA. This work has made use of data from the High Energy
Astrophysics Science Archive Research Center (HEASARC), provided by NASA's
Goddard Space Flight Center.

\def\aj{AJ}%
\def\actaa{Acta Astron.}%
\def\araa{ARA\&A}%
\def\apj{ApJ}%
\def\apjl{ApJ}%
\def\apjs{ApJS}%
\def\ao{Appl.~Opt.}%
\def\apss{Ap\&SS}%
\def\aap{A\&A}%
\def\aapr{A\&A~Rev.}%
\def\aaps{A\&AS}%
\def\azh{AZh}%
\def\baas{BAAS}%
\def\bac{Bull. astr. Inst. Czechosl.}%
\def\caa{Chinese Astron. Astrophys.}%
\def\cjaa{Chinese J. Astron. Astrophys.}%
\def\icarus{Icarus}%
\def\jcap{J. Cosmology Astropart. Phys.}%
\def\jrasc{JRASC}%
\def\mnras{MNRAS}%
\def\memras{MmRAS}%
\def\na{New A}%
\def\nar{New A Rev.}%
\def\pasa{PASA}%
\def\pra{Phys.~Rev.~A}%
\def\prb{Phys.~Rev.~B}%
\def\prc{Phys.~Rev.~C}%
\def\prd{Phys.~Rev.~D}%
\def\pre{Phys.~Rev.~E}%
\def\prl{Phys.~Rev.~Lett.}%
\def\pasp{PASP}%
\def\pasj{PASJ}%
\def\qjras{QJRAS}%
\def\rmxaa{Rev. Mexicana Astron. Astrofis.}%
\def\skytel{S\&T}%
\def\solphys{Sol.~Phys.}%
\def\sovast{Soviet~Ast.}%
\def\ssr{Space~Sci.~Rev.}%
\def\zap{ZAp}%
\def\nat{Nature}%
\def\iaucirc{IAU~Circ.}%
\def\aplett{Astrophys.~Lett.}%
\def\apspr{Astrophys.~Space~Phys.~Res.}%
\def\bain{Bull.~Astron.~Inst.~Netherlands}%
\def\fcp{Fund.~Cosmic~Phys.}%
\def\gca{Geochim.~Cosmochim.~Acta}%
\def\grl{Geophys.~Res.~Lett.}%
\def\jcp{J.~Chem.~Phys.}%
\def\jgr{J.~Geophys.~Res.}%
\def\jqsrt{J.~Quant.~Spec.~Radiat.~Transf.}%
\def\memsai{Mem.~Soc.~Astron.~Italiana}%
\def\nphysa{Nucl.~Phys.~A}%
\def\physrep{Phys.~Rep.}%
\def\physscr{Phys.~Scr}%
\def\planss{Planet.~Space~Sci.}%
\def\procspie{Proc.~SPIE}%
\let\astap=\aap \let\apjlett=\apjl
\let\apjsupp=\apjs 
\bibliographystyle{mn}
\bibliography{mybib}
\end{document}